\documentclass[aps,prfluids,11pt,superscriptaddress,showpacs,floatfix,amsmath,amssymb,longbibliography]{revtex4-1}
\usepackage{color,graphicx,array,tabularx,bm,url,subcaption}
\newcolumntype{x}[1]{>{\centering\arraybackslash\hspace{0pt}}p{#1}}

\begin{document}

\title{Enhancement of drag and mixing in a dilute solution of rodlike polymers
at low Reynolds numbers}

\author{L. Puggioni}
\affiliation{Dipartimento di Fisica and INFN, Universit\`a degli Studi di Torino, via P. Giuria 1, 10125 Torino, Italy.}

\author{G. Boffetta}
\affiliation{Dipartimento di Fisica and INFN, Universit\`a degli Studi di Torino, via P. Giuria 1, 10125 Torino, Italy.}

\author{S. Musacchio}
\thanks{Corresponding author}
\email{stefano.musacchio@unito.it}
\affiliation{Dipartimento di Fisica and INFN, Universit\`a degli Studi di Torino, via P. Giuria 1, 10125 Torino, Italy.}

\begin{abstract}
We study the dynamics of a dilute solution of rigid rodlike polymers in a viscous fluid at low Reynolds number 
by means of numerical simulations of a simple rheological model. We show that the rotational dynamics of polymers 
destabilizes the laminar flow and causes the emergence of a turbulent-like chaotic flow with a wide range of active scales. 
This regime displays an increased flow resistance, corresponding to a reduced mean flow at fixed external forcing, 
as well as an increased mixing efficiency. The latter effect is quantified by measuring the decay of the 
variance of a scalar field transported by the flow. 
By comparing the results of numerical simulations of the model in two- and three-dimensions, 
we show that the phenomena observed are qualitatively independent on the dimensionality of the space. 
\end{abstract}

\date{\today}

\maketitle

\section{Introduction}
\label{sec:intro}
The addition of small amounts of polymers in a fluid causes dramatic effects on
the mechanical properties of the solution.  At high Reynolds number, it is well
known that polymers reduce the turbulent drag compared to that of the solvent
alone \cite{toms1948some}.  The discovery of the phenomenon of drag reduction
has motivated the efforts of the scientific community to investigate the
dynamics of dilute polymer solutions (see, e.g., the reviews
\cite{gyr2013drag,sreenivasan2000onset,white2008mechanics}).  More recently, it
has been discovered that polymer additives alter significantly also the flows
at low Reynolds number.  In this case, even though the Reynolds stresses are
negligible, the elastic stresses give rise to instabilities if the
elasticity of polymers is large enough.  The growth of these instabilities
generate ultimately a chaotic regime which has been called ``elastic
turbulence'' \cite{groisman2000elastic}. 
In this regime, the mixing
efficiency of the flow is strongly enhanced, because the velocity field
develops chaotic structures at small scales, with a power-law energy spectrum
\cite{groisman2001efficient,berti2008two}.  
This phenomenon is extremely useful to increase
the mixing in microfluidic application, where the Reynolds numbers are
typically very low and the diffusive mixing is weak. 

Although most of the studies of these phenomena have been performed with
elastic polymers, these effects can originate also from rigid rodlike polymer.
One advantage of using rodlike polymers in applications is that there are
experimental evidences that the degradation due to large strains is weaker for
rodlike polymers than for elastic polymers \cite{pereira2013drag}.  At large
Reynolds numbers, it has been shown that the drag reduction obtained by
elastic- and rigid-polymers is remarkably similar \cite{virk1997additive,
paschkewitz2004numerical, benzi2005additive, gillissen2008polymer}.
At low Reynolds number, recent numerical studies performed in two-dimensions (2D) have
demonstrated that the addition of rigid polymers originates a chaotic regime
similar to elastic turbulence \cite{emmanuel2017emergence} characterized by
enhanced mixing \cite{musacchio2018enhancement}. 

Here, we extend the investigation of the low-Reynolds chaotic regime in viscous
solutions of rodlike polymers to three dimensional (3D) flows.  We present the
results of numerical simulations in 3D of the rheological model considered in
\cite{emmanuel2017emergence} together with two-dimensional (2D) simulations
for comparison.  
At increasing the concentration of polymers we find an increase of the flow
resistance, quantified by the friction factor, as well as an increased mixing
efficiency. The latter is obtained by measuring the decay rate of the variance
of a scalar field transported in the flow. The comparison of the results of 3D
and 2D simulations reveals that the dimensionality of the space has little
influence on these processes.

\section{Eulerian model for a dilute rods suspension}
\label{sec:mod}

We consider an Eulerian model for a dilute suspension of inertialess rodlike 
polymers with the polymer phase described by the unit-trace 
symmetric tensor field 
$R_{i,j}({\bm x}, t) = \langle n_i n_j \rangle_{\mathcal{V}}$.
$R_{i,j}({\bm x}, t)$ represents the average of the orientation 
vectors ${\bm n}$ 
of individual polymers over an infinitesimal volume element $\mathcal{V}$ 
at position ${\bm x}$ and time $t$ \cite{doi1988theory}.
The polymers are transported by an incompressible velocity field 
${\bm u}({\bm x},t)$. 
The dynamics of the suspension is determined by the following 
coupled equations: 
\begin{subequations}
\begin{alignat}{4}
\partial_t {u_i} + {u_k}\partial_k{u_i} &= -\partial_i p + \nu \partial^2{u_i} + 
\partial_k{\sigma_{ik}} + {f_i}\,,\label{eq:sys1a}
\\
\partial_t R_{ij} + {u_k} \partial_k R_{ij} &= (\partial_k{u_i}) R_{kj}
 + R_{ik}(\partial_k{u_j})-2 R_{ij}(\partial_l u_k) R_{kl},
\label{eq:sys1b}
\end{alignat}
\label{eq1}
\end{subequations}
where $p(\bm x,t)$ is pressure, $\nu$ is the kinematic viscosity of the 
solvent fluid and $\bm f(\bm x,t)$ is the body-force which sustains the flow. 
The form of the polymer stress tensor $\sigma_{ij}$ 
is based on a quadratic approximation proposed by Doi and Edwards 
\cite{doi1988theory}
$\sigma_{ij}=6\nu \eta R_{ij}(\partial_l u_k) R_{kl}$.
The intensity of the polymer feedback on the flow is determined by 
the dimensionless parameter $\eta$ which is proportional to the
polymer concentration.
We remark that model (\ref{eq1}) can also contains additional terms
produced by the orientational diffusion of polymers and which give 
a polymer contribution to the fluid viscosity \cite{doi1988theory}. 
Since we consider dilute concentrations and since, in the presence of 
the flow, the orientation is mainly determined dynamically by the 
velocity gradient, we disregard these terms. 
Numerical simulations at large Reynolds numbers
have shown that model (\ref{eq1}) is able to reproduce the 
main features of turbulent drag reduction in channel flows 
\cite{benzi2005additive,benzi2008comparison,amarouchene2008reynolds}. 

Motivated by previous works which studied the emergence of the chaotic regime 
at low Reynolds number in 2D 
\cite{emmanuel2017emergence,musacchio2018enhancement}, 
here we focus on the case of a 3D viscous bulk flow sustained by 
the Kolmogorov force ${\bm f}({\bm x}) = (F \cos(Kz), 0, 0)$, 
where $F$ is the amplitude and $K$ is the wavenumber of the force. 
In absence of polymers ($\eta=0$) this force produces 
the stationary laminar solution ${\bm u}({\bm x}) = (U_0 \cos(Kz), 0, 0)$ with 
$U_0 = F/(K^2\nu)$, which is linearly stable 
if the Reynolds number $Re=U_0/(\nu K)$ 
is smaller than the critical value 
$Re_c = \sqrt{2}$\cite {meshalkin1961investigation}. 
This flow, which has been first proposed by Kolmogorov as a model to 
understand the transition to turbulence, displays an interesting feature: 
the mean velocity profile $\bar{\bm u}(z)$ 
remains monochromatic even in the turbulent regime,
i.e. $\bar{\bm u}(z) = (U \cos(Kz), 0, 0)$ \cite{musacchio2014turbulent}
(here and in the following the overbar $\bar{[\cdot]}$ 
denotes the average over time $t$ and over the $x$ and $y$ coordinates). 
In analogy with the case of channel flows, the presence of a non-vanishing 
mean velocity profile 
allows us to define the turbulent drag coefficient $f = F/(KU^2)$ 
in terms of the amplitude $U$ of the mean flow, 
which in the turbulent regime is smaller than the laminar solution $U_0$. 
This property has been exploited to study the dependence of the 
turbulent drag in bulk flows on $Re$
\cite{musacchio2014turbulent}
and how it affected by the presence of elastic polymers \cite{boffetta2005drag} 
of rodlike polymers \cite{emmanuel2017emergence}
and inertial particles \cite{sozza2020drag}. 

In the case of dilute suspensions of rodlike polymers 
described by the equations (\ref{eq1}) with $\eta >0$,
the Kolmogorov forcing produces the laminar flow 
$(U_0 \cos(Kz), 0, 0)$ with amplitude $U_0 = F/(K^2\nu)$ 
independent on the polymer concentration. 
This is at variance with viscoelastic models in which the presence of 
polymers affects the amplitude of the laminar flow \cite{boffetta2005drag}.
The steady solution of the equation (\ref{eq:sys1b}) 
for the conformation tensor requires $R_{i3} = R_{3i} = 0$ and $ \partial_x R_{ij} =0$, 
that is, polymers are oriented in the $x-y$ plane and their orientation 
depends only on the $y$ and $z$ coordinates. 

\section{Numerical Results}
\label{sec:res}

We performed a set of numerical simulations of (\ref{eq1}) 
discretized on a regular grid of $N^3 = 256^3$ gridpoints, 
on a triply periodic domain of size $L=2\pi$. 
The time integration uses a fourth-order Runge-Kutta scheme 
with implicit integration of the linear dissipative terms. 
In all the simulations the viscosity is set to $\nu=1$ 
and the flow is sustained by the Kolmogorov force 
${\bm f}({\bm x}) = (F \cos(Kz), 0, 0)$, 
with forcing wavenumber $K=4$ 
and forcing amplitude $F=\nu^2K^3$, 
such that, in absence of polymers ($\eta=0$),
the flow is laminar with Reynolds number $Re= 1 < Re_c$. 
The feedback coefficient is varied from $\eta =5$ to 
$\eta =8$. Experimentally this corresponds, for the
case of an aqueous solution of xanthan gum, to 
concentrations in the range of $73-102$ 
wppm \cite{amarouchene2008reynolds}. 
A diffusive term $\kappa \partial^2 R_{ij}$ with $\kappa = 4 \times 10^{-3}$ 
is added to eq. ~(\ref{eq:sys1b}) 
in order to improve the numerical stability \cite{sureshkumar1995effect}. 

\begin{figure}[h!]
\includegraphics[width=0.6\textwidth]{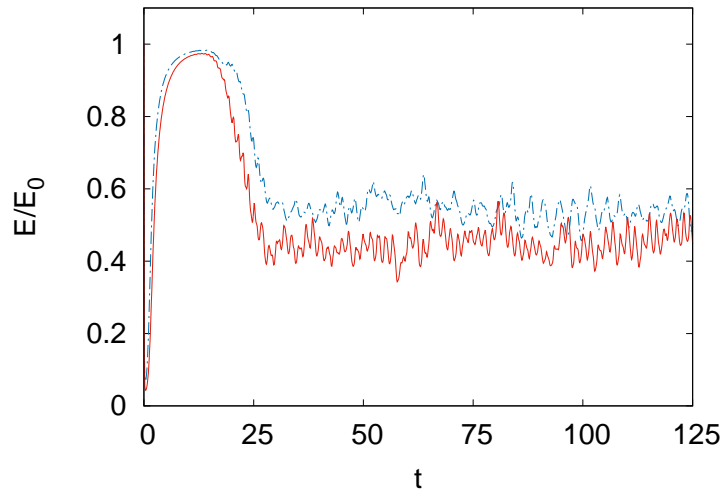}
\caption{Kinetic energy $E$, normalized with the laminar energy $E_0$, 
for two simulations in 3D with $\eta = 6$ (blue dashed line) 
and $\eta = 8$ (red solid line).}
\label{fig1}
\end{figure}

At time $t=0$ we initialize the velocity field to the fixed-point laminar 
solution while the components of the tensor $R$ are initially distributed 
randomly with isotropic distribution. 
Note that the latter is not a steady solution of the Equation~(\ref{eq:sys1b}). 
The typical evolution of the system is well illustrated by the time evolution of the 
the kinetic energy $E = \frac{1}{2} \langle | {\bm u}|^2 \rangle $, 
shown in Figure~\ref{fig1} for two simulations with $\eta = 6$ and $\eta = 8$
(here and in the following $\langle \cdot \rangle$ denotes the average over the
whole volume).  
After an initial transient the laminar flow destabilizes 
and eventually reaches a statistically stationary chaotic state  
with considerably less kinetic energy than that of the laminar flow 
$E_0 = \frac{1}{2} U_0^2$. 
In this regime, the kinetic energy displays rapid oscillations  
whose frequency increases with the polymer concentration, 
while the average value of $E$ decreases at increasing $\eta$.
\begin{figure}[h!]
\includegraphics[width=0.28\linewidth]{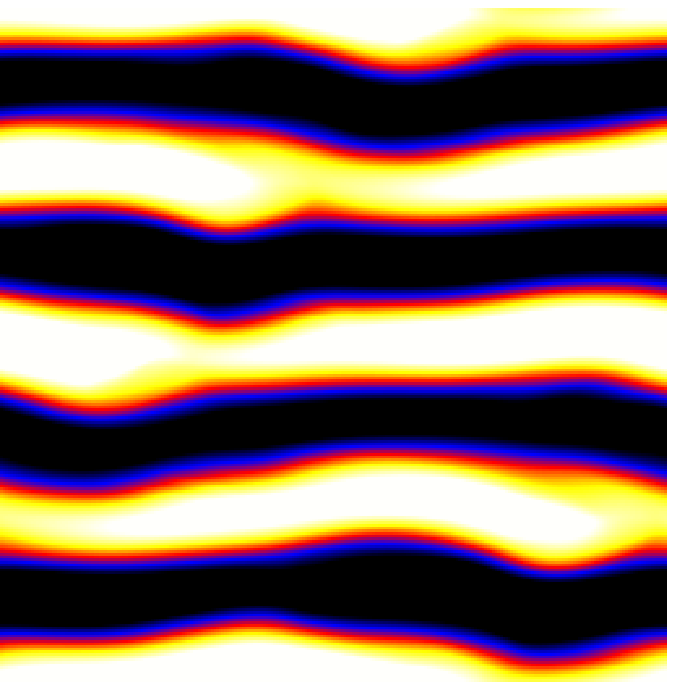}
\includegraphics[width=0.28\linewidth]{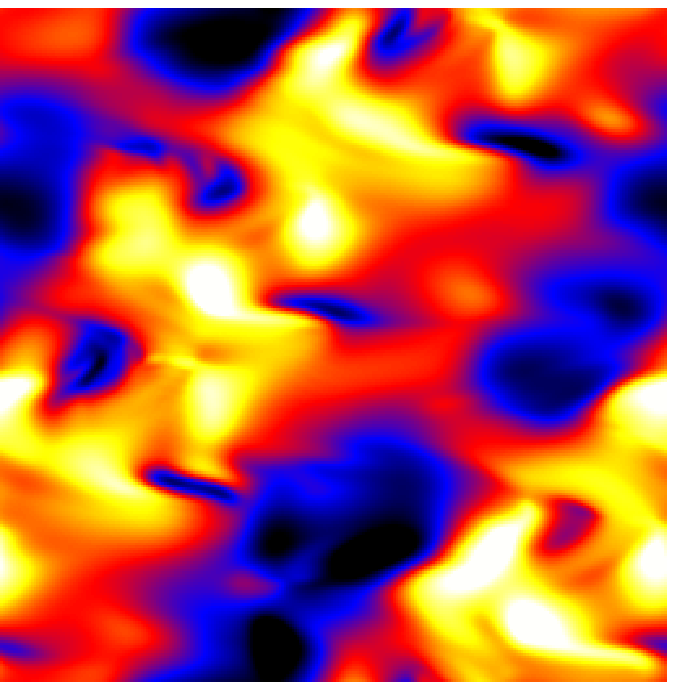}
\includegraphics[width=0.28\linewidth]{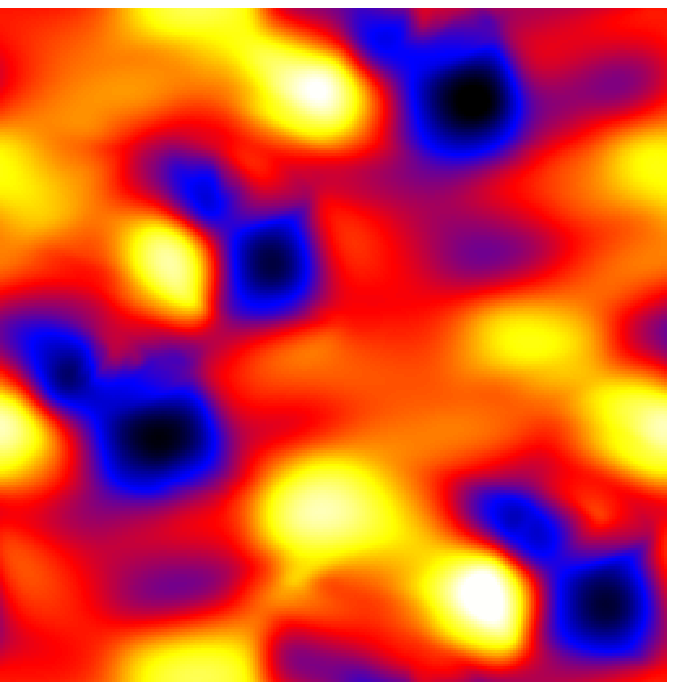}
\caption{Vertical sections in the $x-z$ plane of the velocity components 
$u_x$, $u_y$, $u_z$ (from left to right) in the 3D chaotic regime for 
$\eta = 7$. The color scale ranges from $-3 u'_i$ (black) to $3u'_i$ (white),  
where $u'_i$ are the rms values of the velocity fluctuations.}
\label{fig2}
\end{figure}

We have studied the influence of the initial conditions by performing
three simulations with different realizations of the initial random 
configuration of the conformation tensor. 
The ensembles obtained from different initial conditions are 
statistically equivalent in the chaotic regime. 
Conversely, the initial transient is dependent on the details of the initial configuration. 
In the case of initial random orientation of polymers, 
at the very beginning we observe a rapid alignment of the polymers 
in the direction of the mean flow which causes a strong dissipation of energy. 
After that, the external force restores the laminar flow and the kinetic energy grows up to value close to $E_0$. 
Finally, the laminar flow destabilizes and the system reaches the chaotic state. 
 
In Figure \ref{fig2} we show three sections of the velocity components 
$u_x$,$u_y$ and $u_z$ in the plane $x-z$ at fixed $y$ from the simulation 
with $\eta=7$ in the chaotic regime. 
While the structure of the Kolmogorov flow remains present in the $u_x$ field, 
it also displays irregular motion at small-scales. 
The form of these structures is  clearly visible in the $u_y$ and $u_z$ fields in which the mean flow is absent.
They qualitatively resemble the elastic waves observed in viscoelastic flows 
\cite{berti2010elastic,varshney2019elastic}.

\subsection{Statistics of the velocity}
\begin{figure}[h!]
\includegraphics[width=0.6\textwidth]{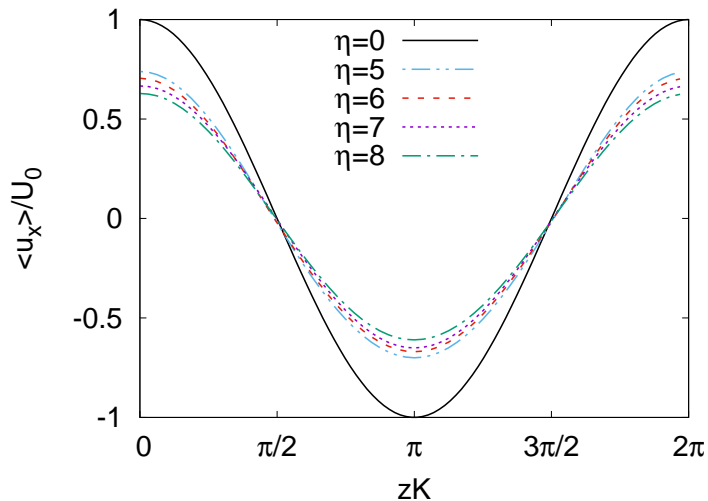}
\caption{Profiles of the mean velocity $\langle u_x(z) \rangle$ 
averaged over $x$, $y$ and time, 
in 3D simulations with different values of $\eta$.
The solid black line corresponds to the laminar solution
of the Newtonian fluid at $\eta=0$.}
\label{fig3}
\end{figure}

One relevant feature of the Kolmogorov flow is that, also in the
chaotic and in the turbulent regimes, it maintains a monochromatic mean 
flow $\langle u_x \rangle = U \cos (z/L)$. This feature is
true even in the presence of polymers, as shown in fig. \ref{fig3}
where we plot the average mean profile for different concentrations.
We observe that the amplitude of the mean flow is reduced with respect
to the laminar solution, as a consequence of the chaotic motion induced
by polymers.

It is therefore natural to decompose the velocity field in a mean 
(monochromatic) component and fluctuations as
\begin{equation}
\textbf{\textit{u}} \left( \textbf{\textit{x}} \right) = 
U \left( \cos \left(z/L \right), 0, 0 \right) + 
\textbf{\textit{u}}'(\textbf{\textit{x}})
\label{eq3}
\end{equation}
Table~\ref{table1} reports the values of the rms velocity fluctuations 
together with the amplitude of the mean flow and other relevant 
quantities. 

\begin{table}[h!]
\begin{tabular}{cccccccccc}
$\eta$ & U & S & $\Sigma$ & $u'_x$ & $u'_y$ & $u'_z$ & $\varepsilon_I$ & $\varepsilon_{\nu}$ & $\varepsilon_{p}$ \\ \hline
5 \qquad & \quad 2.87 & \quad 0.10 & \quad 4.40 & \quad 0.64 & \quad 0.12 & \quad 0.40 & \quad 91.9 & \quad 74.8 & \quad 17.1 \\
6 \qquad & \quad 2.74 & \quad 0.10 & \quad 5.02 & \quad 0.63 & \quad 0.13 & \quad 0.39 & \quad 87.2 & \quad 68.6 & \quad 18.6  \\
7 \qquad & \quad 2.63 & \quad 0.10 & \quad 5.57 & \quad 0.64 & \quad 0.16 & \quad 0.39 & \quad 83.2 & \quad 63.6 & \quad 19.6 \\
8 \qquad & \quad 2.48 & \quad 0.09 & \quad 6.08 & \quad 0.69 & \quad 0.18 & \quad 0.40 & \quad 78.8 & \quad 58.3 & \quad 20.6 \\
\end{tabular}
\caption{Parameters of the 3D simulations. $U$ is the amplitude of the mean
velocity, $S$ the amplitude of the Reynolds stress and $\Sigma$ that of the
polymer stress.  $u'_x$, $u'_y$ and $u'_z$ are the rms of the three components
of velocity fluctuations. $\varepsilon_I$ is the mean energy input,
$\varepsilon_{\nu}$ the viscous energy dissipation and $\varepsilon_{p}$ 
the mean polymer dissipation.}
\label{table1}
\end{table}

\begin{figure}[h!]
\includegraphics[width=0.6\linewidth]{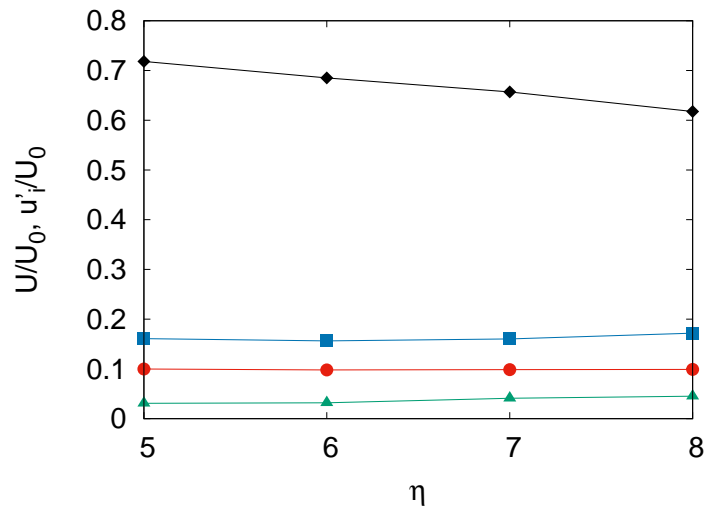}
\caption{Mean velocity profiles $U$ (black diamonds)
  and components of rms velocity fluctuations
  ($u'_x$ blue squares, $u'_y$ green triangles, $u'_z$ red circles)
  in 3D simulations with different values of $\eta$.}
\label{fig4}
\end{figure}

The amplitudes of the velocity components, averaged over time, are shown 
in Fig.~\ref{fig4}. 
Figure~\ref{fig4}a shows that the amplitude of the mean flow is 
significantly reduced with respect to the laminar case and that this 
effect is stronger at increasing values of the concentration parameter $\eta$. 
The rms values of velocity fluctuations are, on the contrary, weakly 
dependent on $\eta$. We notice that fluctuations along streawise direction 
$u'_x$ are dominant followed by fluctuation in the $z$ direction, 
while fluctuations in the spanwise direction $u'_y$ are much smaller. 
This result suggest that, even in the chaotic regime, the flow remains 
approximatively two-dimensional. 

\subsection{Drag and momentum budget}
\begin{figure}[h!]
\includegraphics[width=0.6\textwidth]{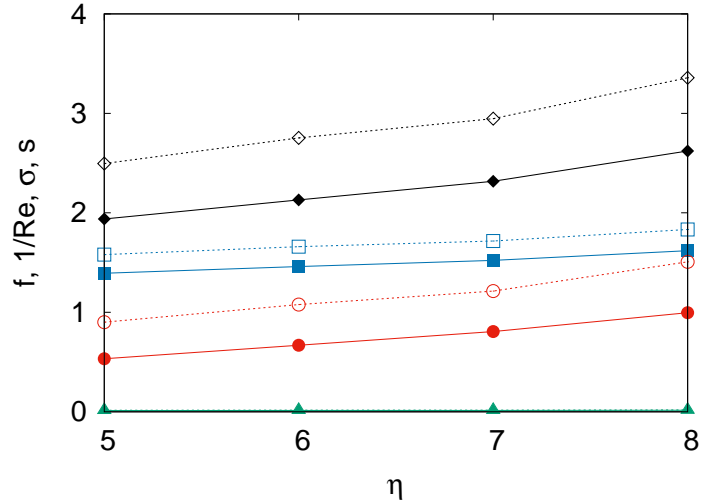}
\caption{
Friction factor $f$ (black diamonds)
normalized viscous stress $1/Re$ (blue squares), 
polymer stress coefficient $\sigma$ (red circles) 
and Reynolds stress coefficient $s$ (green triangles), as
function of $\eta$. 
Filled symbols are for the three-dimensional DNS, empty ones are for the
two-dimensional DNS.}
\label{fig5}
\end{figure}

In order to better understand the effect of polymers on the mean flow we
consider the momentum budget. Averaging (\ref{eq1}) in stationary
conditions over $x$, $y$ and $t$, we obtain the stress budget
\begin{equation}
\partial_z \Pi_r = \partial_z \left( \Pi_{\nu} + \Pi_p \right) + f_z,
\label{eq4}
\end{equation}
where $\Pi_r = \overline{u_x u_z}$ is the Reynolds stress, 
$\Pi_{\nu} = \nu \partial_z \overline{u_x}$ the viscous stress, 
and $\Pi_p = \overline{\sigma_{xz}}$ the polymer stress. 
In the steady state all these quantities have a monochromatic profile
\begin{equation}
\Pi_r = S \sin (K z), \qquad \Pi_{\nu} = -\nu K U \sin (K z), \qquad \Pi_p = -\Sigma \sin (K z), 
\label{eq5}
\end{equation}
and therefore (\ref{eq4}) becomes an algebraic equation for the coefficients
\begin{equation}
S K + \nu K^2 U + \Sigma K = F .
\label{eq6}
\end{equation}
The dimensionless version of the momentum budget is obtained by dividing all the 
terms of (\ref{eq6}) by $K U^2$ and defining the friction coefficient 
$f=F/(K U^2)$, which quantify the ratio between the work done by the force and the
kinetic energy of the mean flow, 
the Reynolds stress coefficient $s=S/U^2$ and 
the polymer stress coefficient $\sigma=\Sigma/U^2$:
\begin{equation}
f = {1 \over Re} + s + \sigma
\label{eq7}
\end{equation}
Figure ~\ref{fig5} shows that increasing the concentration of polymers
produces an increase of the friction factor
which is mostly due to the increase of the polymer stress and  
partly to a weaker increase of the viscous stress. 
The Reynolds stress remains negligible. 

\begin{figure}[h!]
\includegraphics[width=0.6\textwidth]{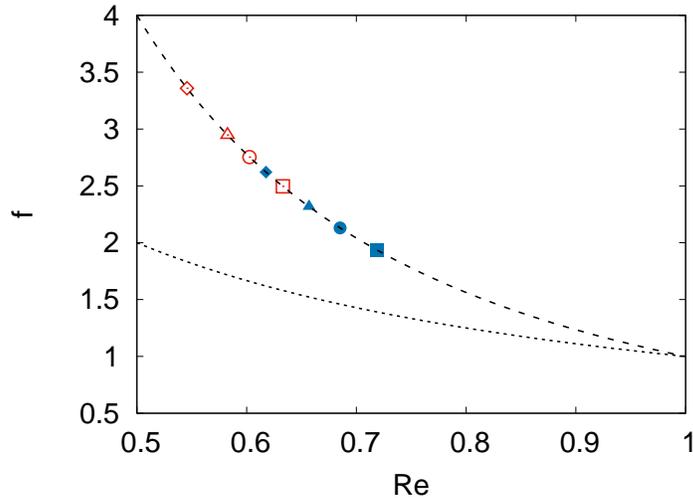}
\caption{Friction factor $f$ as a function of $Re$ 
in the 3D (blue, filled simbols) and 2D (red, empty symbols) simulations 
with different values of $\eta$: 
$\eta=5$ (squares), 
$\eta=5$ (circles), 
$\eta=6$ (triangles), 
$\eta=7$ (diamonds).
The dotted line is the laminar law $f = 1/Re$,
while the dashed line is $f=Re_0/Re^2$.}
\label{fig6}
\end{figure}

By definition, the drag coefficient $f$ is linked to the Reynolds number 
by $f = Re_0/Re^2$ where $Re_0 = U_0 /K\nu = F / K^3 \nu^2$. 
The effects of polymers is therefore twofold: 
They reduces the Reynolds number of the flow and increases its resistance. 
Note that the contribution of the viscous stress to the increase of the drag coefficient 
is subdominant ($\propto 1/Re$) with respect to that of the polymer stress ($\propto 1/Re^2$). 
This is clearly shown in Figure ~\ref{fig6} in which the friction factor $f$ 
is plotted as a function of $Re$ for the  different values of $\eta$.
Since both $f$ and $Re$ do not depend explicitely on $\eta$, 
points corresponding to simulations at the same $F$ and $\nu$ lays 
on the line $Re_0/Re^2$. The intersection in Figure ~\ref{fig6} 
of this line with the laminar
line $1/Re$ corresponds to the laminar flow in the absence of polymers 
($\eta=0$).

\subsection{Energy budget}

Additional information regarding the effects of polymer on the flow is 
obtained by the analysis of the energy budget. 
By multiplying (\ref{eq1}) by $\textbf{\textit{u}}$ and integrating 
over the volume we get the balance equation for the mean kinetic
energy (we note that, unlike the case of elastic polymers, we cannot 
associate a deformation energy to rigid polymers)
\begin{equation}
\frac{d}{dt} \langle E \rangle = 
\varepsilon_I - \varepsilon_{\nu} - \varepsilon_{p},
\label{eq8}
\end{equation}
where 
$\varepsilon_I = \langle \textbf{\textit{f}} \cdot \textbf{\textit{u}} \rangle = FU/2$ 
is the mean energy input, 
$\varepsilon_{\nu}= \langle \nu \left| \nabla \textbf{\textit{u}}\right|^2 \rangle$ 
the mean viscous dissipation, and
$\varepsilon_{p} = \langle \sigma_{ij} \partial_j u_i \rangle$ 
is an additional dissipation of kinetic energy 
due to the coupling with polymers.
We remark that the local value of the term 
$\sigma_{ij} \partial_j u_i$ can be either positive or negative, meaning
that polymers can locally either give or subtract energy from the flow.
Nonetheless the volume average is always negative, indicating that the 
global effect of polymer is dissipative.
This is due to the fact that the coupling between
the rods and the fluid arise from viscous forces at molecular scale, 
whose mean effect is to dissipate a part of the kinetic energy
~\cite{doi1988theory}.
 
\begin{figure}[h!]
\includegraphics[width=0.6\textwidth]{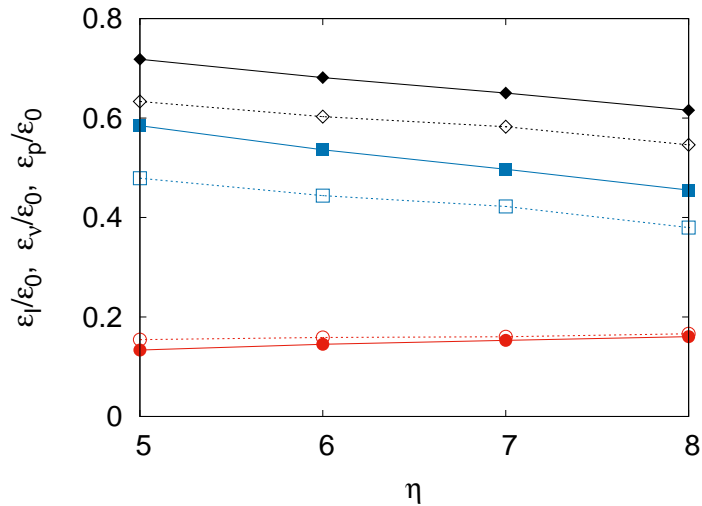}
\caption{Mean values of energy input $\varepsilon_I$ (black diamonds),
  viscous dissipation $\varepsilon_{\nu}$ (blue squares)
  and polymer dissipation $\varepsilon_{p}$ (red circles),
  as function of $\eta$.
  Filled symbols are for the three-dimensional DNS, empty ones are for the two-dimensional DNS.}
\label{fig7}
\end{figure}

In the statistical steady state, averaging over sufficiently
long times, the energy can be considered constant, and therefore
we have the energy balance
$\varepsilon_I = \varepsilon_{\nu} + \varepsilon_{p}$. 
These quantities are shown in Fig.~\ref{fig7},
normalized with the mean energy input of the laminar flow 
$\varepsilon_0 = FU_0/2$. 
We observe a slight increase in the polymer dissipation as the 
concentration coefficient increases, but the main effect of polymer
is a suppression of the energy input provided by the constant forcing
as a consequence of the reduction of the mean flow amplitude. 
This is consistent with the results plotted in Fig.~\ref{fig2} 
showing that kinetic energy is reduced at increasing polymer concentration.
Figure~\ref{fig7} indicates that for all values of $\eta$, energy
is mostly dissipated by viscosity. Therefore we expect that the 
small scale properties of the flow are weakly affected by the polymer
concentration. 

\begin{figure}[h!]
\includegraphics[width=0.6\textwidth]{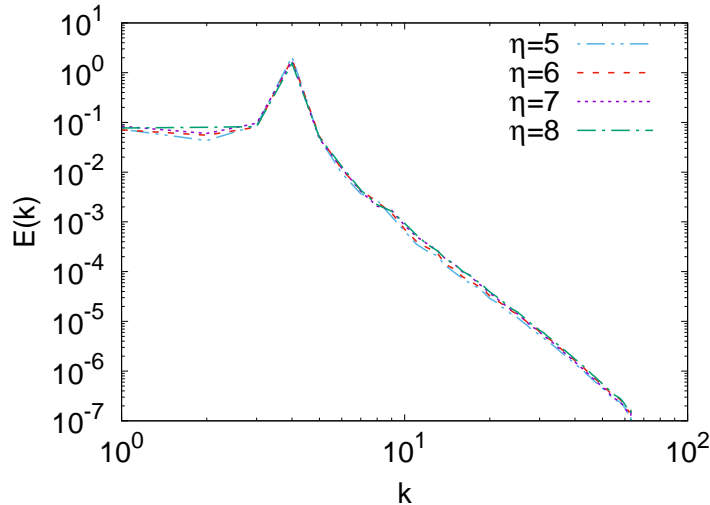}
\caption{Energy spectra, averaged on time, obtained in 3D simulations with 
different values of $\eta$.}
\label{fig8}
\end{figure}

In order to investigate this point, in Fig.~\ref{fig8} we plot the kinetic
energy spectra in stationary conditions and for the different values of 
concentrations. 
Note the peak of the spectra at the forcing wavenumber $K=4$.
We observe very small variations of the spectrum with
$\eta$, mostly concentrated at small wavenumbers (since the total energy
changes with $\eta$). At large wavenumbers the spectra display a power
law behavior $E \left( k \right) \sim k^{-\alpha}$ with $\alpha \simeq 4.8$,
an indication of the presence of fluctuations at all scales.
The fact that the power spectrum is steeper than $k^{-3}$ 
indicates that velocity field is smooth in this regime, similarly to 
what observed in elastic turbulence \cite{groisman2000elastic}.

\subsection{Mixing properties}

The presence of velocities fluctuations over a wide range of spatial scales
has a strong influence on the mixing efficiency of the flow. 
In order to address this issue we have integrated the equation
for a scalar field $\theta({\bm x},t)$ transported
by the velocity field ${\bm u}$ obtained from Eqs.~(\ref{eq1}): 
\begin{equation} 
\partial_t \theta + {u_k}\partial_k \theta = D \partial^2 \theta 
\label{eq9}
\end{equation}
where $D$ is the molecular diffusivity, 
which is set to $D= 2 \cdot 10^{-3}$ in the simulations. 
The integration of the Equation~(\ref{eq9}) is started
at an arbitrary time $t_0$ in the stationary regime of chaotic flow. 
We chose a monochromatic initial condition for $\theta$,
with the same periodicity of the mean flow $\theta({\bm x},t_0) = \cos(K z)$.  
With this initial condition, in absence of polymers,
the mixing is due exclusively to molecular diffusion
because the gradient of the scalar field $\nabla \theta$ is orthogonal 
to the laminar velocity field.  
In particular, for $\eta=0$ the variance of the scalar field 
(as well as the variance of its gradient) 
decays exponentially as
$\langle \theta^2 \rangle
\propto  \langle (\nabla \theta)^2 \rangle
\propto \exp \left( -\beta_0 (t-t_0) \right)$, with $\beta_0 = 2DK^2$. 

\begin{figure}[h!]
\includegraphics[width=0.28\linewidth]{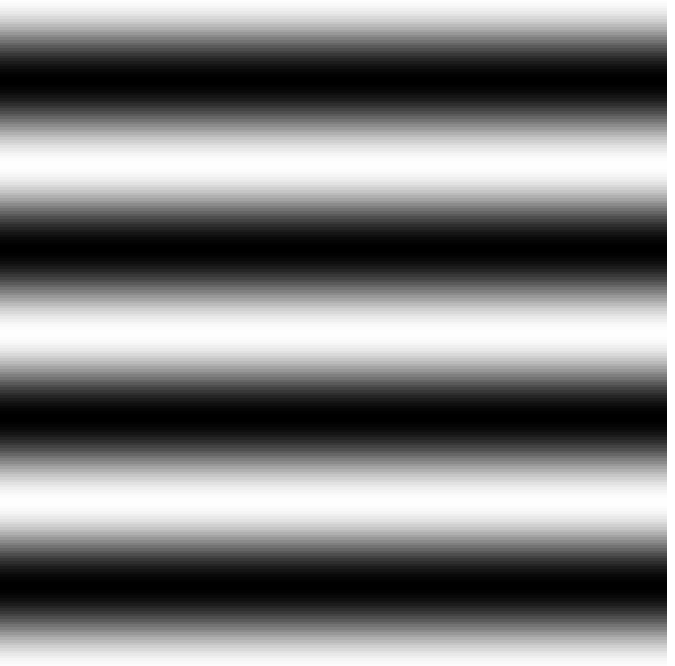}
\includegraphics[width=0.28\linewidth]{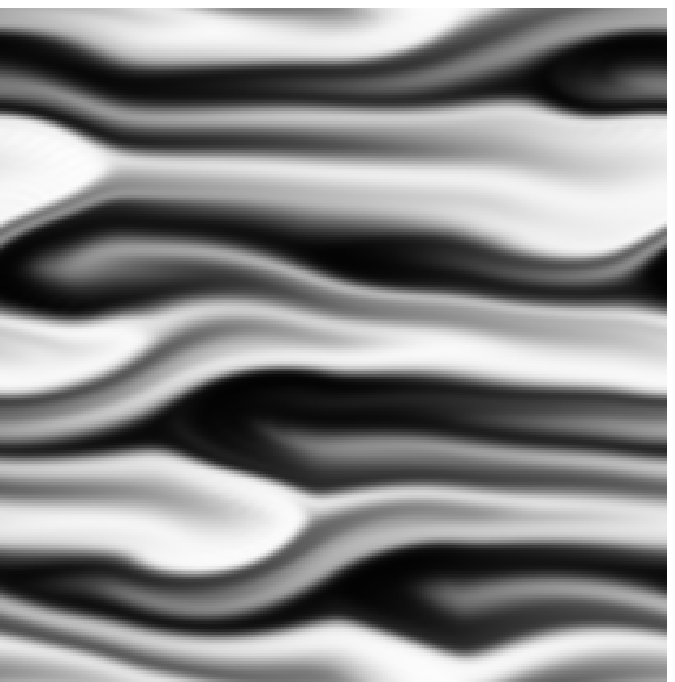}
\includegraphics[width=0.28\linewidth]{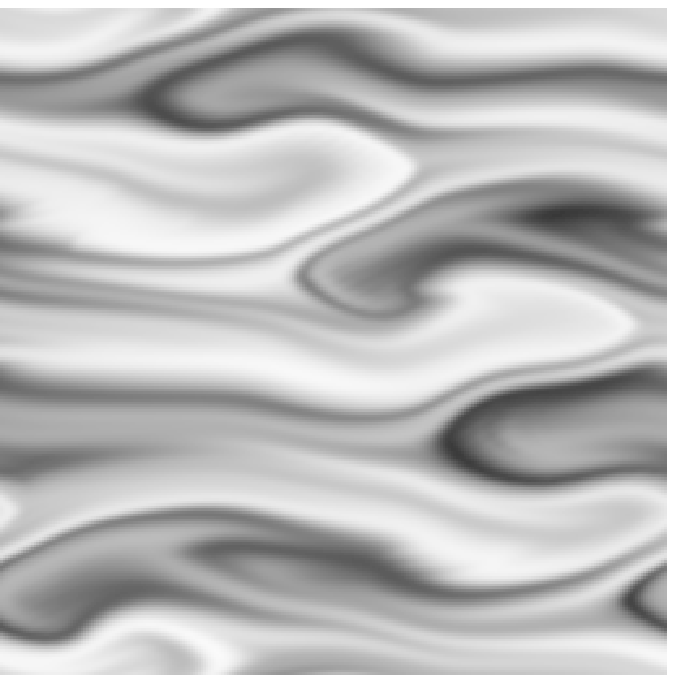}
\caption{Vertical section in the $x - z$ plane at fixed $y=0$
          of the scalar field $\theta$ for the 3D simulation with $\eta=8$ 
          at different times. 
	  From left to right: $t-t_0=0$, $t-t_0=2$, $t-t_0=4$.} 
\label{fig9}
\end{figure}

\begin{figure}[h!]
\includegraphics[width=0.48\linewidth]{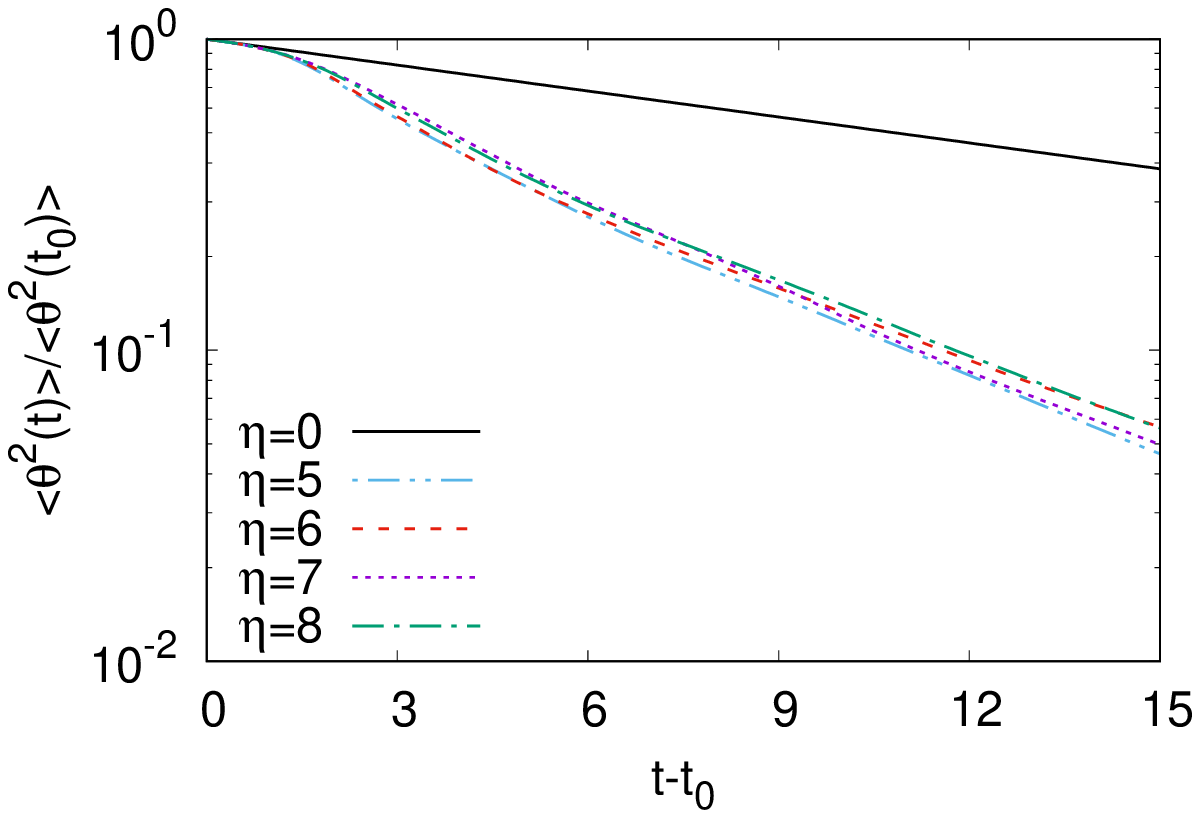}
\includegraphics[width=0.48\linewidth]{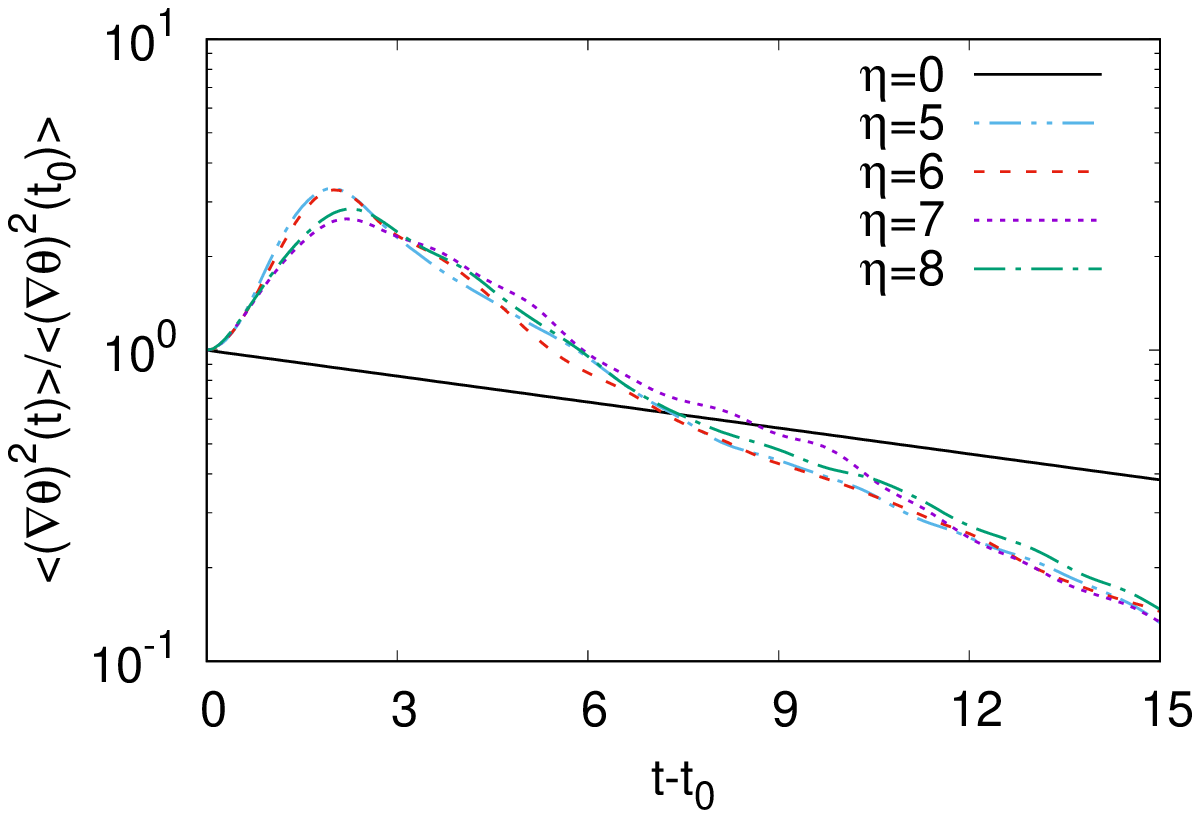}	
\captionof{figure}{Decay of the variance of the scalar field $\langle \theta^2 \rangle$  (left panel) 
and of the scalar gradients $\langle\nabla \theta)^2 \rangle$ (right panel)
for different values of $\eta$ in 3D simulations.
Solid black line represents the diffusive exponential decay in absence of polymers.}
\label{fig10}
\end{figure}

In presence of polymers we observe a strong enhancement of the mixing, 
which is illustrated by the vertical sections of $\theta$ shown in Fig.\ref{fig9}.
At variance with the diffusive case at $\eta=0$
in which the scalar field remains monochromatic,
here we observe the formation of thin scalar filaments,
which rapidly transfer the scalar fluctuations to small dissipative scales.
This process enhances the mixing efficiency. 
A quantitative measure of the mixing is provided by the temporal evolution
of the variance of $\theta$ and $\nabla \theta$ shown in Figure~\ref{fig10}.
Here and in the following, the results presented have been averaged
over $13$ independent simulations of Equation~(\ref{eq9}),
starting from the same initial condition
$\theta({\bm x},t_0)$
at different times $t_0$, i.e., with different velocity fields. 
The decay of $\langle \theta^2 \rangle$ in the chaotic flow induced by the polymers
is faster with respect to the case $\eta=0$.
The same result is observed for the long-time decay of the
variance of scalar gradients $\langle (\nabla \theta)^2 \rangle$, 
even though the chaotic advection of the scalar field causes an initial increase
of its gradients. 
For the concentration values considered in our study,
we do not observe a strong dependence of the mixing efficiency on $\eta$. 

The instantaneous exponential decay rate of the scalar variance 
$\beta_{p}=-{d \over dt} \log \langle \theta^2 \rangle$
can be written, using (\ref{eq9}), as
\begin{equation}
\beta_p \left( t \right) = -\frac{d}{dt} \log \langle \theta^2 \rangle = 
2 D  \frac{\langle (\nabla \theta)^2\rangle}{\langle \theta^2 \rangle} 
\end{equation}
which can be directly compared with the decay rate
due to molecular diffusion $\beta_0 = 2DK^2$.

\begin{figure}[h!]
\includegraphics[width=0.6\textwidth]{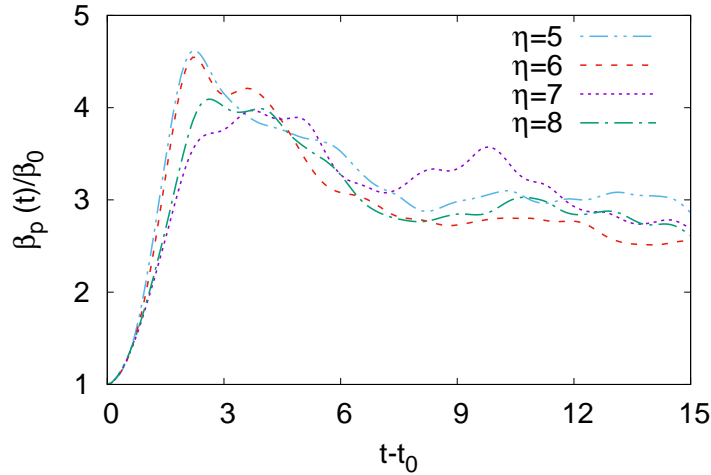}
\caption{Instantaneous exponential decay rate
          $\beta_p \left( t \right)$ for different values of $\eta$
          in 3D simulations.}
\label{fig11}
\end{figure}

The decay rate $\beta_p$ reaches a maximum value after a very short time,
corresponding to the maximum development of thin filaments of the scalar field. 
At long time, since both $\langle \theta^2 \rangle$ and
$\langle (\nabla \theta)^2\rangle$ decay exponentially,
$\beta_p$ approaches an almost constant value,
about three times larger than $\beta_0$ (see Fig.~\ref{fig11})
which quantifies the increased mixing efficiency.
We note that the ratio $\beta_p/\beta_0$ is proportional 
to the square of the ratio between the large scale of the scalar field $1/K$ 
and the typical scale of its gradient
$(\langle \theta^2 \rangle/ \langle (\nabla \theta)^2 \rangle)^{1/2}$. 

\subsection{Comparison between 2D and 3D}

The results presented so far show that 
the properties of the chaotic flow which is obtained 
from 3D numerical simulations of the model (\ref{eq1}) 
for a dilute solution of rigid rods 
are qualitatively similar to those reported in previous numerical studies in 2D
\cite{emmanuel2017emergence,musacchio2018enhancement}. 
In particular, we have found that the fluctuations of the $y$-component of the velocity $u_y$,
which is transverse both to the streamwise direction of the mean flow $x$ 
and to the direction of its gradient $z$,  
are considerably smaller than those of $u_x$ and $u_z$ 
(see Figure \ref{fig4}). 
This confirms that the dynamics of the three-dimensional system 
is substantially two-dimensional, 
and therefore the properties of the chaotic flow 
are qualitatively independent on its dimensionality. 

\begin{figure}[h!]
\includegraphics[width=0.6\textwidth]{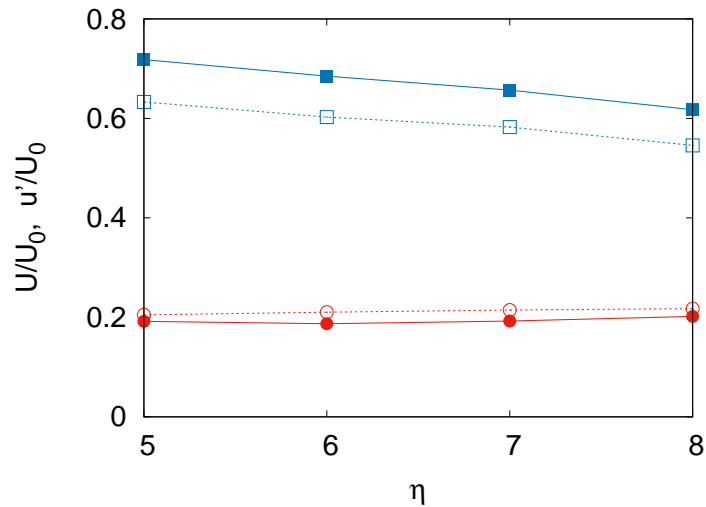}
\caption{Amplitudes of mean velocity profiles $U$ (blue squares) and
		rms velocity fluctuations $u'_{rms}$ (red circles) as a function of $\eta$
		in 3D (filled symbol) and 2D (empty symbols) simulations.}
\label{fig12}
\end{figure}

In order to compare quantitatively the properties of the 3D and 2D flows, 
we have performed a set of 2D simulations of the system of equations (\ref{eq1}) 
with the same parameters of the 3D simulations. 
The comparison of the mean flow and velocity fluctuations reported in Figure~\ref{fig12}
shows that the effects of polymer are more pronounced in 2D than in 3D.
At fixed value of the polymer concentration $\eta$,  
we find that the velocity fluctuations are more intense in 2D than in 3D. 
Similarly, the reduction the amplitude $U$ of the mean flow with respect to the laminar solution $U_0$ 
is stronger in 2D than in 3D. 
It is worth to notice that 2D and 3D curves of $U$ and $u'$ as a function of $\eta$ are almost parallel, 
which indicates that the effects of dimensionality is systematic and it is independent on $\eta$. 

The comparison of the momentum balance is reported in Figure \ref{fig5}. 
Also in this case we observe that the values of the friction factor in 2D 
are systematically higher than in 3D at fixed $\eta$. 
In both cases, the drag enhancement is mostly due to the increase of the polymer stress,
with a subdominant contribution due to the increase of the viscous stress. 
The combined effect of increased friction factor and reduced Reynolds number 
is clearly visible in Figure \ref{fig6}, 
in which the deviation from the Newtonian point $f=Re=Re_0$ 
is stronger for the 2D simulations. 
In the energy balance, the reduction of the amplitude of the mean flow 
causes a reduction of the energy injection rate $\varepsilon_I$ 
in 2D simulations with respect to the 3D ones at fixed $\eta$ (see  Figure \ref{fig7}).  
This phenomenon is balanced by a reduction of the viscous dissipation rate $\varepsilon_\nu$, 
while the energy dissipation due to polymers remains almost unchanged. 

In summary, we can conclude that the effects of rod-like polymers in viscous flows 
in three-dimensions is weaker than in two-dimensions. 
The origin of this difference is probably due to the different rotational degrees of freedom of the rods. 
In 2D, the rotation of the polymers can occur only in the $x-z$ plane. 
This imply that, during the rotation, the $R_{33}$ component of the conformation tensor is non-zero 
and therefore the polymers are oriented in the direction of the gradient of the mean flow (the $z$-direction). 
Conversely, in 3D they can rotate also in the $x-y$ plane. 
Indeed we have observed that in the stationary regime the average values of $R_{22}$ and $R_{33}$ are very similar.
The consequences of polymer rotations in the $x-z$ and $x-y$ planes 
on the polymer stress tensor $\sigma_{ij}$ are very different. 
We remind that $\sigma_{ij}$ is proportional to the product of the configuration tensor $R_{ij}$ 
and the velocity gradient tensor $\partial_i u_j$. 
In the case of the laminar Kolmogorov flow ${\bm u}({\bm x}) = (U_0 \cos(Kz), 0, 0)$, 
the only component of the velocity gradient which is non-zero is $\partial_3 u_1$.
As a consequence, there is no stress induced by rotations in the $x-y$ plane (allowed only in 3D). 
In the case of the chaotic flow, the gradients of velocity in the $y$-direction 
originates only from the fluctuating part of the velocity field, 
therefore they are significantly smaller than those in the $z$-direction.  
As a result, the polymer stress in 3D is on average weaker than in 2D flow with the same $\eta$.

\begin{figure}[h!]
\includegraphics[width=0.6\textwidth]{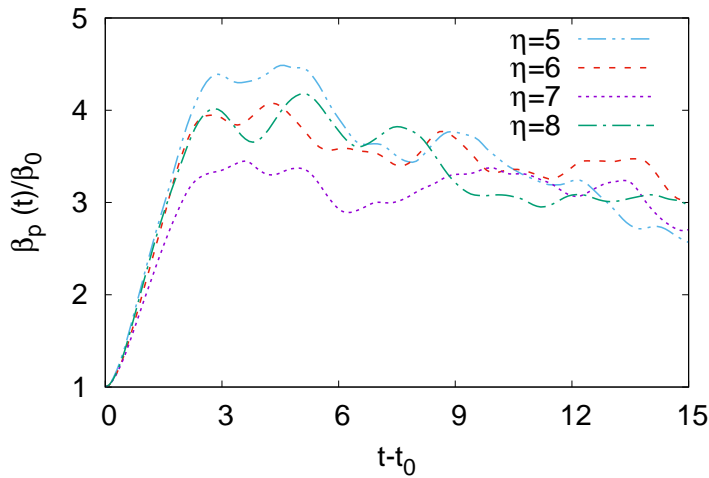}
\caption{Instantaneous exponential decay rate
          $\beta_p \left( t \right)$ for different values of $\eta$
          in 2D simulations.}
\label{fig13}
\end{figure}
Finally we compare the mixing properties of 2D and 3D flows by integrating the transport equation (\ref{eq9}) 
for a two-dimensional $\theta$ scalar field transported by the 2D flow. Initial condition and diffusion coefficient 
are identical to those of 3D simulations. 
The values of the instantaneous exponential decay rate $\beta_p \left( t \right)$ 
obtained in the 2D simulations are shown in Figure~\ref{fig13}. 
They are very similar to those obtained in 3D simulations. 
This is in agreement with the observation that the intensity of velocity fluctuations, 
which causes the mixing, is very similar as well (see Fig.~\ref{fig12}). 

\section{Conclusions}
\label{sec:conclus}

We studied the dynamics of rigid rodlike polymer solutions at 
low Reynolds numbers by means of direct numerical simulations
of a rheological model both in 2D and in 3D.
We have found that the presence of polymers induces a chaotic, turbulent-like
flow with increased flow resistance and enhanced mixing efficiency at 
Reynolds numbers at which the laminar solution for the Newtonian fluid 
without polymers is linearly stable.
The phenomenology observed is qualitatively independent on the 
dimensionality, but we found that, for the same values of the parameters,
the effects are stronger in the 2D case. This difference 
is explained in terms of the different rotational degrees of freedom of the rods. 

Future numerical work at higher resolution and/or with a different
numerical scheme would allow to reach more realistic values of the 
parameters. This will also open the possibility to compare the numerical
simulations with the results obtained from laboratory experiments. 



\section*{Acknowledgments}
We acknowledge HPC CINECA for computing resources
(INFN-CINECA grant no. INFN21-FieldTurb).
We acknowledges support from the Departments of Excellence grant (MIUR).



%

\end{document}